\documentclass[prb,twocolumn,showpacs,preprintnumbers,amsmath,amssymb]{revtex4}

\usepackage[dvips]{graphicx}
\usepackage{dcolumn}
\usepackage{bm}
\usepackage{tabularx}

\newcolumntype{L}{>{\raggedright\arraybackslash}X}
\newcolumntype{C}{>{\centering\arraybackslash}X}
\newcolumntype{R}{>{\raggedleft\arraybackslash}X}

\begin{document}


\title{The Mixed State of Charge-Density-Wave \\
in a Ring-Shaped Single Crystals}

\author{Masahiko Hayashi}
 \email{hayashi@cmt.is.tohoku.ac.jp}
 \author{Hiromichi Ebisawa}
\affiliation{%
Graduate School of 
Information Sciences, Tohoku University, 
Aramaki Aoba-ku, Sendai 980-8579, Japan and \\
JST-CREST, 4-1-8 Honcho, Kawaguchi, Saitama 332-0012, Japan
}%
\author{Kazuhiro Kuboki}
\affiliation{Department of Physics, Kobe University, Kobe 657-8501, Japan}

\date{\today}

\begin{abstract}
Charge-density-wave (CDW) phase transition in a ring-shaped 
crystals, recently synthesized by Tanda et al. 
[Nature, {\bf 417}, 397 (2002)], is studied 
based on a mean-field-approximation of 
Ginzburg-Landau free energy. 
It is shown that in a ring-shaped crystals 
CDW undergoes frustration 
due to the curvature (bending) 
of the ring (geometrical frustration) 
and, thus, forms a mixed state analogous to what 
a type-II superconductor forms under a magnetic field. 
We discuss the nature of the phase transition in 
the ring-CDW in relation to recent experiments. 
\end{abstract}

\pacs{61.72.Bb,61.82.Rx,71.45.Lr,74.25.Op
}
\maketitle

\section{Introduction}

Charge-density-wave (CDW) has been attracting intense interest 
from solid state physicists, because of its 
unique physical properties as an electronic condensate
\cite{Gruner,Gorkov}. 
Especially analogy to superconductivity 
has been stimulating imagination of various 
people. 
Fr\"olich paid attention to collective 
transport (sliding) of CDW and 
investigated within it the origin of superconductivity 
before the foundation of BCS theory 
\cite{Froelich}. 
Although it has been clarified that the 
sliding motion of CDW is easily suppressed by 
inhomogeneities, 
realization of the Fr\"olich's superconductivity in 
ultraclean CDW crystals is an intriguing 
target of experimental trails. 

Similarities and dissimilarities between vortices in superconductors and 
dislocations in CDW systems are also interesting topics 
of this field. 
Analogy between the phase slip phenomena via vortex ring 
in a superconducting wire and the dislocation nucleation 
in a sliding CDW has been pointed out
\cite{Ong-Verma-Maki}. 
Several people attempted to search the counterpart of the  
mixed state of type-II superconductors in 
CDW systems 
\cite{Brazovsky,Hayashi-Yoshioka,Hayashi-Yoshioka2}. 
According to Ref. \onlinecite{Hayashi-Yoshioka}, 
CDW systems can also form a mixed state analogous 
to that in type-II superconductors 
when an electric field perpendicular to the chains is applied. 
This \lq\lq mixed state\rq\rq\, of CDW, however, is 
formed only in the surface region of the sample 
within a depth of Debye screening length, 
which is usually several lattice spacings in most CDW materials. 
Therefore direct experimental observation of this state is considered to be very difficult. 

Recently Tanda et al. have succeeded in synthesizing a 
ring crystal of CDW materials\cite{1d}, where the conducting chains 
are along the azimuthal direction\cite{Tanda,Okajima}. 
In a ring crystal, CDW system feels 
frustration due to bending, 
just like a superconductor feels frustration under a magnetic field. 
Thus we can expect that the CDW forms a mixed state 
as in a type-II superconductor. 
This fact was first pointed out by Nogawa and Nemoto 
\cite{Nogawa} and the dynamics of ring-CDW is investigated 
in terms of the Monte Carlo simulation on a lattice model of CDW. 
In their analysis, ring CDW is mapped to a uniformly frustrated XY model. 

Experimental investigation of the ring-CDW's has been 
carried out by Shimatake et al. \cite{Shimatake}.  
They measured the single particle relaxation rate 
in ring- and whisker-CDW's using femtosecond spectroscopy and 
found that quasiparticle density in 
ring CDW's shows a stretched exponential decay
$\propto \exp\{(t/\tau)^{0.3}\}$, whereas 
it decays exponentially in whisker samples. 
Moreover relaxation time in ring CDW's did not show 
significant divergence at $T_c$, 
which is common in whisker samples. 
Indication of such slow dynamics in ring CDW's is also 
found in the simulations by 
Nogawa and Nemoto\cite{Nogawa}, 
who found a power law decay of relaxation time 
in ring crystals. 
Although details of relaxation formula are different, 
both results suggest the existence of slow relaxation 
processes in ring CDW's, 
which may be due to the dynamics of dislocations 
intrinsic to ring CDW's. 
In this context it is important to study how dislocations 
intrinsic to ring CDW's is distributed in the ordered state. 

In this paper we first derive the Ginzburg-Landau (GL)
free energy of CDW in a 
ring geometry. 
Our results are slightly different from
uniformly frustrated XY model, or 
superconductors in a magnetic field, especially 
when the radial thickess of the ring is large compared 
to the radius. 
We then study the mean field behavior of the system 
in order to clarify the transition temperature of CDW 
in a ring geometry. 
We also comment on the analogy to superconductors 
in a magnetic field, which is useful for thin rings 
(radial thickness $\ll$ radius). 
The relation between our results and some experimental results 
are also discussed. 

\section{Ginzburg-Landau free energy of a ring CDW}

In constructing the free energy describing ring CDW's 
we need detailed knowledge about the crystal structure of the 
ring crystals. 
We refer to 
Ref.\onlinecite{Hayashi-Ebisawa-Kuboki_ring}
on this point and summarize the results. 
According to this reference, crystal defects 
(edge dislocations) are almost unifromly distributed 
in ring crystals with diameters $\gtrsim 1\mu$m. 
This means that the chains forming the ring are 
almost free from elastic stress (stretch or contraction) and 
can be described by free CDW chains with different lengths 
depending on the radius. 
There may be some stretch or contraction of 
the order of lattice spacing for each chain 
due to the discrete nature of the lattice, however, 
this is negligible 
for rings with realistic diameters $\sim 1\mu$m 
which is much larger than the unit cell of the lattice. 

Then we start from the following GL free energy, 
\begin{align}
F=\sum_{k=1}^{N_z}\left\{
\sum_{j=1}^N F^{\rm c}_j + \sum_{j=1}^{N-1} F^{\rm int}_{j,j+1}
\right\},
\label{free_energy}
\end{align}
where the first and the second term correspond to 
intra-chain and inter-chain contribution, respectively. 
The number of the chains is denoted by $N$. 
Summation over $k$ corresponds to the stacking 
of the layer along the axis of the ring, 
with $N_z$ being the number of the layers. 
Since we assume that the system is uniform along the axis, 
$k$-summation is suppressed hereafter. 
Here $F^{\rm c}_j$ is 
the free energy of the $j$-th chain given by 
\begin{align}
F^{\rm c}_j = \int_0^{2\pi R_j}d x\,&
\biggl[\xi_0^2
\left|\partial_x \psi_j(x) \right|^2 +
\nonumber\\
&+(t-1)|\psi_j(x)|^2+
\frac{1}{2}|\psi_j(x)|^4
\biggr],
\label{free_energy2}
\end{align}
where $R_j$ is the radius of the $j$-th chain given by 
$R_j=R+d \left(j-\frac{N}{2}\right)$
($R$ is the radius at the center of the width 
and $d$ is the inter-chain spacing), 
$\xi_0$ is the coherence length of CDW at 
zero temperature $T=0$, and 
$t$ is the reduced temperature $T/T_c$ with 
$T_c$ being the transition temperature in the bulk. 
$\psi_j(x)$ is the order parameter on the $j$-th chain. 

\begin{figure}[htb]
\includegraphics[width=6cm,clip]{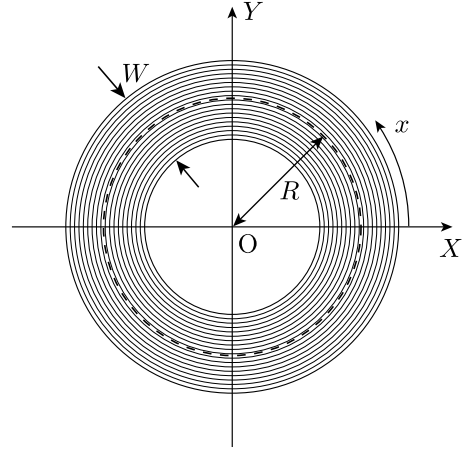}
    \caption{
	Geometry and parameters of our system. 
	The $z$-axis is perpendicular to the sheet. 
          }
    \label{geometry}
\end{figure}

The inter-chain coupling term is 
constructed in the following way: 
We write the order parameter as  $\psi_j \equiv |\psi_j|e^{i \phi_j(x)}$. 
Then the local charge density modulation due to CDW 
can be written as $\delta\rho^{\rm local}_j(x)
\propto  |\psi_j(x)| \cos \{2 k_F x+\phi_j(x)\}$, 
and the inter-chain coupling is given in the following form, 
\begin{align}
&F^{\rm int}_{j,j+1} = -
\frac{R_j+R_{j+1}}{2} 
\int_0^{2 \pi} d\theta\, J_0 |\psi_j (R_j \theta)||\psi_{j+1}(R_{j+1} \theta)|
\nonumber\\
&\times\cos \{2 k_F R_j \theta+\phi_j(x)\}
\cos \{2 k_F R_{j+1} \theta+\phi_{j+1}(x)\}
\label{ic_couple}
\end{align}
where $J_0$ is a constant. 
CDW usually prefers $\pi$-phase shift between 
neighboring chains due to, for example, 
Coulomb repulsion\cite{Gruner}. 
We have beforehand taken it into account 
in the definition of the order parameter, 
and the minus sign of Eq. (\ref{ic_couple}) is 
due to this. 
Noting the equality, 
\begin{align}
&\cos \{2 k_F R_j \theta+\phi_j(x)\}
\cos \{2 k_F R_{j+1} \theta+\phi_{j+1}(x)\}
\nonumber\\
&=\frac{1}{2}\cos \{2 k_F (R_j+R_{j+1}) \theta+\phi_j(x)+\phi_{j+1}(x)\}
\nonumber\\
&\phantom{=}+\frac{1}{2}\cos \{2 k_F d \theta+\phi_j(x)-\phi_{j+1}(x)\},
\label{cos}
\end{align}
where $d = R_{j+1} - R_j$, 
and neglecting the first term of the right hand side
as the negligible fast oscillation, 
we obtain, 
\begin{align}
&F^{\rm int}_{j,j+1} \simeq-
\frac{R_j+R_{j+1}}{2} 
\int_0^{2 \pi} d\theta\, J_0 |\psi_j (R_j \theta)||\psi_{j+1}(R_{j+1} \theta)|
\nonumber\\
&\phantom{aaaaa}
\times\frac{1}{2}\cos \{2 k_F d \theta+\phi_j(x)-\phi_{j+1}(x)\}
\nonumber\\
& = \frac{R_j+R_{j+1}}{2} \frac{J_0}{4} 
\int_0^{2 \pi} d\theta\, 
\nonumber\\
&\phantom{aaaaa}\times
|\psi_j (R_j \theta)e^{i k_F d \theta}-
\psi_{j+1}(R_{j+1} \theta)e^{-i k_F d \theta}|^2. 
\label{free_energy3}
\end{align}
In the last line, we added some terms proportional to $|\psi_j|^2$, 
in order to subtract the effect of the renormalization of $T_c$.

Next we gauge transform the order parameter as 
$\psi_j(x) \rightarrow \psi_j(x)e^{i 2 k_F d \theta j}$ 
and obtain the resulting free energy as, 
\begin{align}
F^{\rm c}_j = &R_j\int_0^{2\pi}d \theta\,
\biggl[\xi_0^2
\left|\left(\frac{1}{R_j}
\frac{\partial}{\partial\theta}- i \frac{2 k_F d j}{R_j}
\right) \psi_j(R_j\theta) \right|^2 +
\nonumber\\
&+(t-1)|\psi_j(R_j\theta)|^2+
\frac{1}{2}|\psi_j(R_j\theta)|^4
\biggr],
\label{free_energy_g1}
\end{align}
and \begin{align}
&F^{\rm int}_{j,j+1} =
\frac{R_j+R_{j+1}}{2}\frac{J_0}{4}
\int_0^{2 \pi} d\theta\, 
|\psi_j (R_j \theta)-
\psi_{j+1}(R_{j+1} \theta)|^2. 
\label{free_energy_g2}
\end{align}

Since we are interested in the 
behavior of CDW near $T_c$, we have disregarded the 
effects of electrostatic screening. 
In order words, the scalar potential does not appear in our 
equations. 
This is an over simplification when we go to 
lower temperatures, where CDW order is well developed 
and charge modulation is significant. 
This point will be discussed further in the end of this paper. 

The boundary condition for $\psi_j(x)$ in the ring-geometry 
is given by 
\begin{align}
&\psi_j(0)=\psi(R_j),
\label{bc1} \\
&\partial_x\psi_j (0) = \partial_x\psi_j (R_j). 
\end{align}
These conditions impose 
$\phi_j(R_j)-\phi_j(0) = - 4 \pi k_F R_j+2 \pi N$, 
where $N$ is an integer.  
This gives rise to the phenomena 
similar to the Little-Parks effect in superconducting rings. 
For systems much larger than the 
Fermi wave length, $R_j \gg 1/k_F$, 
this effect is very weak and negligible. 
This applies for most ring CDW systems synthesized until now. 

Next we consider a change of coordinate 
(from chain to chain) 
$x \rightarrow x'$, so that 
$R_j$ becomes $R$ for all the chains 
(Namely, $x/R_j = x'/R$). 
Here defining $\chi(R_j)\equiv 2 (R_j - R) k_F/R$, we obtain 
\begin{align}
F^{\rm c}_j =\frac{R_j}{R}
&\int_0^{2 \pi R} dx'\,
\biggl[
\left(
\frac{R}{R_j}
\right)^2\xi_0^2
\left|
\left(
\partial_{x'} + i \chi(R_j)
\right)
\psi'_j(x')
\right|^2\nonumber\\
&+(t-1)|\psi'_j(x')|^2
+\frac{1}{2}|\psi'_j(x')|^4
\biggr].
\label{disc_f1}
\end{align}
The inter-chain term can be rewritten as 
\begin{align}
F^{\rm int}_{j,j+1} &= \frac{R_j+R_{j+1}}{2 R} 
\int_0^{2 \pi R} dx'\, \frac{J_0}{4} |\psi'_j (x') - \psi'_{j+1}(x')|^2. 
\label{disc_f2}
\end{align}

Now we apply continuum approximation to 
the radial direction ($y$-direction). 
Putting $W=N d$, we obtain 
\begin{align}
F^{c}
= 
&\frac{1}{d}\int_{R-W/2}^{R+W/2}dy \left(
\frac{y}{R}
\right)\int_0^{2 \pi R}
dx'\,
\nonumber\\
\biggl[&\left(
\frac{R}{y}
\right)^2\xi_0^2
\left|
\left(
\partial_{x'} + i \chi(y)
\right)
\psi'(x',y)
\right|^2\nonumber\\
&+(t-1)|\psi'(x',y)|^2
+\frac{1}{2}|\psi'(x',y)|^4
\biggr],
\end{align}
and 
\begin{align}
F^{\rm int}&= \frac{1}{d}\int_{R-W/2}^{R+W/2}dy
\left(\frac{y}{R}\right)
\int_0^{2 \pi R} dx'\, \frac{J_0}{4} d^2 |\partial_y \psi' (x',y)|^2. 
\end{align}
It is worthwhile to point out that 
the free energy $F=F^{\rm c} +F^{\rm int}$ has 
rather complicated $y$ dependences due to 
the bending. 
It is not simply the type of the frustrated XY model, 
or a superconductor in a magnetic field. 
Especially, the factor $(y/R)$ appearing before 
$x'$-integral comes from the fact that inner chains have 
smaller lengths. 
This can make the thermodynamic properties of ring CDW's 
different from those in superconductors in a magnetic field 
especially when the ring is thick $W \sim R$. 
This point should be carefully studied in further analysis. 

\section{Case of a Thin Ring $W \ll R$: 
A superconducting analogy}

Although thick rings ($W \sim R$) are a little different from 
superconductors in a magnetic field, as we have seen above, 
the analogy to superconductors is still useful for thin rings 
($W \ll R$). 
For thin rings, the 
free energy is simplified to 
\begin{align}
F=
\frac{1}{d}&
\int_0^{2 \pi R} dx'\,
\int_{R-W/2}^{R+W/2}d y\, 
\nonumber\\
\biggl[&
\xi_0^2
\left\{\left|
\left(
\partial_{x'} + i \chi(y)
\right)
\psi(x',y)
\right|^2
+\gamma^2  
|\partial_y \psi(x',y)|^2
\right\}
\nonumber\\
&+(t-1)|\psi(x',y)|^2
+\frac{1}{2}|\psi(x',y)|^4
\biggr],
\end{align}
where $\gamma$ is defined by $J_0/4=\gamma^2 \xi_0^2 /d^2$. 
This is actually the continuum version of the lattice model 
previously introduced by Nogawa and Nemoto
\cite{Nogawa}. 

We further change the coordinates as $y'=y/\gamma$ 
so that the system is isotropic in both $x$- and 
$y$-direction and obtain 
\begin{align}
F=
\frac{\gamma}{d}
\int_0^{2 \pi R} dx'\,&
\int_{(R-W/2)/\gamma}^{(R+W/2)/\gamma}d y'\, 
\nonumber\\
\biggl[
\xi_0^2
&\left|
\left(
\nabla' + i \vec{\chi}(\vec{r'})
\right)
\psi(\vec{r'})
\right|^2
\nonumber\\
&+(t-1)|\psi(\vec{r'})|^2
+\frac{1}{2}|\psi(\vec{r'})|^4
\biggr], 
\label{iso_free_energy}
\end{align}
where $\vec{r'} = (x',y')$ and 
$\displaystyle \vec{\chi}(\vec{r'})=
2\left(\frac{\gamma y'}{R}-1\right)k_F\vec{e}_x$ 
with $\vec{e}_x$ being the unit vector in $x$-direction. 

At this stage, we see a nice analogy to the free energy of 
a superconductor in a magnetic field. 
The magnetic field $H$ corresponds to the 
curvature of the ring $1/R$ as 
\begin{align}
\frac{2 \pi H}{\phi_0} \Longleftrightarrow \frac{2 k_F \gamma}{R}. 
\label{analogy}
\end{align}
From this analogy we can deduce the transition temperature of 
CDW in a ring crystal. 
We note that a ring CDW corresponds to a type-II superconductor, 
not type-I, 
since there is no mechanism to screen the \lq\lq curvature\rq\rq\, 
(or bending) 
of the ring in the CDW system.  
In other words, there is no Meissner phase. 


Since the transition temperature of a type-II 
superdconductor under a magnetic field $H$ is given by 
\begin{align}
T_{c2}(H) &=  T_c(0)\left(1 - \frac{H}{H_{c2}(0)}\right),
\end{align}
where $H_{c2}(0)=\phi_0/(2 \pi \xi_0^2)$ is the 
upper critical field at $T=0$
(Note that this formula is valid if $H \ll H_{c2}(0)$.
In that sense, $H_{c2}(0)$ is the critical field at $T=0$ 
obtained from the linear extrapolation from $T\sim T_c$.), 
the CDW transition temperature of a ring crystal of radius $R$ 
is given by 
\begin{align}
T_{c2}(R) =  T_c(\infty)
\left(1 - \frac{R_{c2}}{R}\right), 
\label{rc2}
\end{align}
where 
$R_{c2}=2 k_F \gamma \xi_0^2$ 
and $T_c(\infty)$ is the transition temperature of 
the whisker sample. 

Here we apply our results to actual NbSe$_3$ 
ring crystal. 
We put 
$2 k_F \simeq 4.3$nm$^{-1}$, 
$\xi_0\simeq 6.1$nm and 
$\gamma \simeq 1/10\sim 1/100$ 
for NbSe$_3$, and then 
we obtain 
$R_{c2}\simeq 160\times \gamma\,$nm $= 1.6 \sim 16$nm. 
\\

\section{Numerical Study of Linearized Ginzburg-Landau Equation}

We study the linearized GL equation for the cases of arbitrary 
$R$ and $W$ and determine the CDW transition temperature. 
We start by 
disregarding the quartic term of the order parameter 
in Eqs. (\ref{disc_f1}). 
By introducing the Fourier transformation, 
\begin{align}
\psi_j (x) = \frac{1}{\sqrt{2 \pi R}}\sum_{n=-\infty}^\infty
{\tilde\psi}_j (n) e^{i n x/R},
\end{align} 
the free energy $F$ is rewritten as 
\begin{align}
&F = \sum_{n=-\infty}^\infty \sum_{j = 1}^N 
\biggl[\frac{R}{R_j}\xi_0^2\left(\frac{n}{R}+\chi(R_j)\right)^2
\nonumber\\
&\phantom{aaaaaaaaaaaaaaa}
+\frac{R_j}{R}(t-1)\biggr] |{\tilde\psi}_j(n)|^2
\nonumber\\
&+\sum_{n=-\infty}^\infty \sum_{j = 1}^{N-1}
\frac{\gamma^2\xi_0^2}{d^2}\frac{R_j+R_{j+1}}{2 R}
\left|{\tilde\psi}_j(n)-
{\tilde\psi}_{j+1}(n)\right|^2.
\end{align}
We put ${\tilde \Psi}_j(n) \equiv \sqrt{R_j/R} \,{\tilde\psi}_j(n)$, 
and the above equation becomes 
\begin{align}
&F = \sum_{n=-\infty}^\infty \sum_{j = 1}^N 
\biggl[
\left(\frac{R}{R_j}\right)^2
\xi_0^2\left(\frac{n}{R}+\chi(R_j)\right)^2
+(t-1)\biggr] |{\tilde\Psi}_j(n)|^2
\nonumber\\
&+\sum_{n=-\infty}^\infty \sum_{j = 1}^{N-1}
\frac{\gamma^2\xi_0^2}{d^2}\frac{R_j+R_{j+1}}{2}
\left|\frac{{\tilde\Psi}_j(n)}{\sqrt{R_j}}-
\frac{{\tilde\Psi}_{j+1}(n)}{\sqrt{R_{j+1}}}\right|^2
\nonumber\\
&\equiv\sum_{n=-\infty}^\infty \sum_{j,k = 1}^N 
{\tilde \Psi}_j^* (n)
\cdot \left\{M_{jk}(n)+(t-1)\delta_{jk}\right\} \cdot
{\tilde\Psi}_k(n).
\end{align}
The integer $n$ indexes the mode of the order parameter. 
All the eigenvalues of $M_{jk}(n)$ are positive 
and we denote the smallest one by $\lambda_{\rm min}(n)$. 
The mode indexed by $n$  becomes unstable 
when $\lambda_{\rm min}(n) + t-1 < 0$ is satisfied. 
Below this temperature 
${\tilde \Psi}_j(n)$ (or ${\tilde \psi}_j(n)$) tends to have a nonzero expectation value 
and the nonlinear term of GL free energy becomes relevant. 
The mode which becomes unstable at 
the highest temperature dominates the real CDW transition. 
Therefore, by diagonalizing $M_{jk}(n)$ and comparing the 
eigenvalues of all possible $n$'s, we can estimate 
the CDW transition temperature. 

\begin{figure}[htb]
\includegraphics[width=7cm,clip]{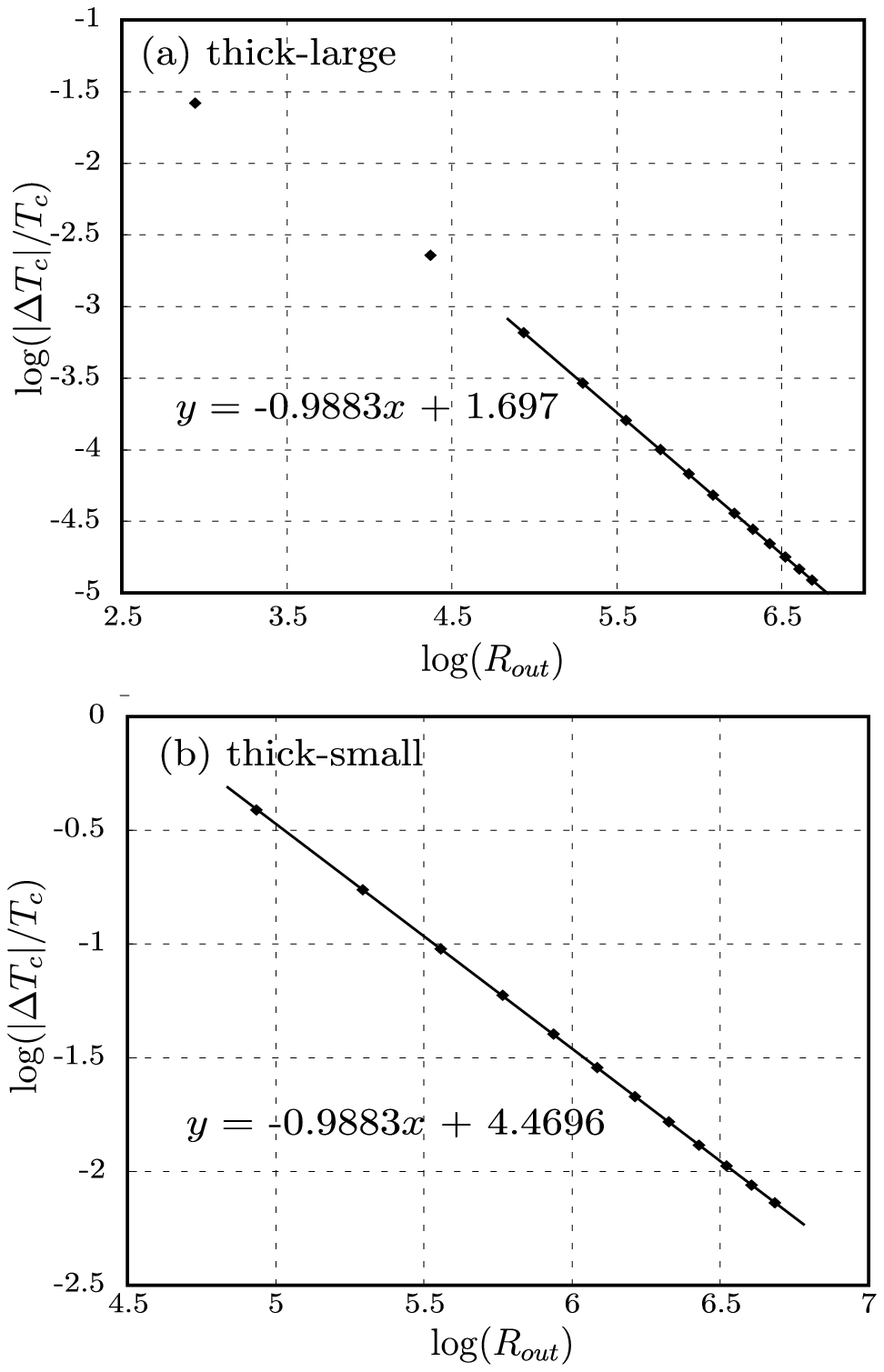}
    \caption[test]{
   Log-log plot of the outermost radius $R_{out}$ vs.
    the shift of the transition temperature in thick ($W \sim 2 R$) rings 
    (or cylinders). 
    (a) $\xi_0 = 5 d$, (b) $\xi_0 = 20 d$. 
    Lines are the fitting line to the linear part of the data. 
    $R_{out}$ is in the unit of $d$. 
          }
    \label{tc_thick}
\end{figure}

\begin{figure}[htb]
\includegraphics[width=7cm,clip]{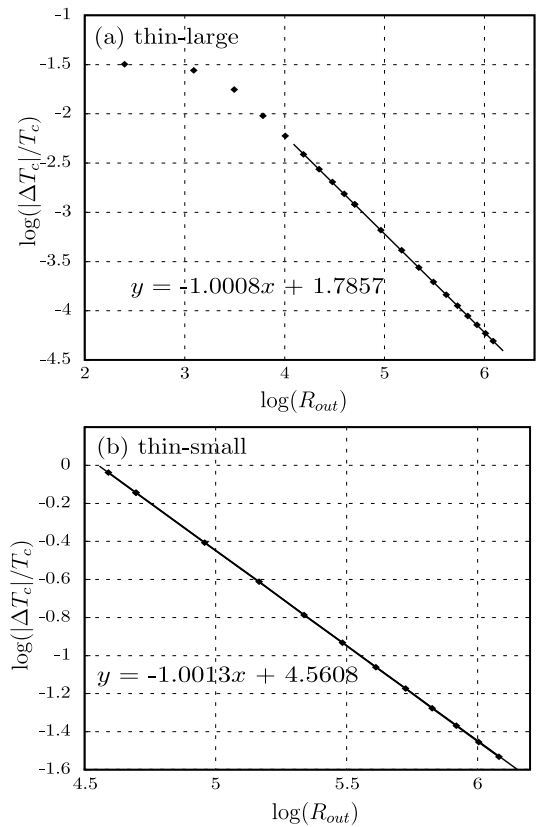}
    \caption[test]{
    Log-log plot of the outermost radius $R_{out}$ vs.
    the shift of the
     transition temperature in thin ($W = R/5$) rings. 
    (a) $\xi_0 = 5 d$, (b) $\xi_0 = 20 d$. 
    Lines are the fitting line to the linear part of the data. 
    $R_{out}$ is in the unit of $d$. 
     }
    \label{tc_thin}
\end{figure}

In Figs. \ref{tc_thick} and \ref{tc_thin}, 
we have shown the log-log (natural logarithm) plot of 
the shift of the transition temperatures of several systems 
as a function of the radius of outermost chain $R_{out}
=R+W/2$.  
($\Delta T_c$ is the decrease of the transition temperature from 
the bulk $T_c$.)
Fig. \ref{tc_thick} is for thick samples where $W \sim 2 R$ 
($W = 2 (R-d)$ to be more precise). 
In this case, the ring looks almost like a disk. 
Fig. \ref{tc_thin} is for thin samples where we put $W = R/5$. 
(a) and (b) in each figure are data for different coherence length $\xi_0$. 
(a) is for $\xi_0 = 5 d$ and (b) is for $\xi_0 = 20 d$. 
From these figures, we can see that amount of $\Delta T_c$ 
is inversely proportional to $R_{out}$ for samples with larger diameters. 

\begin{figure}[htb]
\includegraphics[width=7cm,clip]{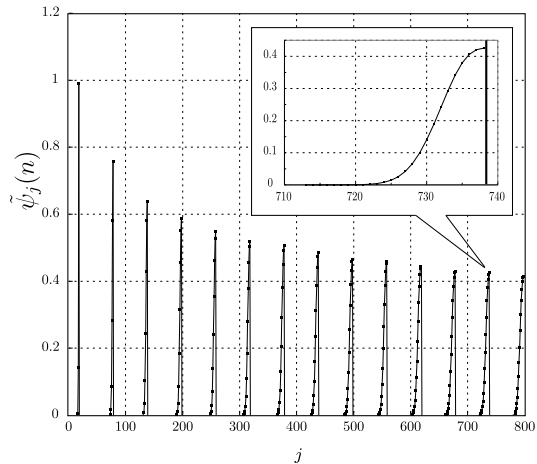}
    \caption[test]{
    The amplitude of the order parameter 
    at the CDW transition temperature 
    for samples with several different radiuses. 
    Each peak corresponds to solutions for different 
    radius. 
    The peak is always located at the outer edge of the sample. 
    The inset is the magnification of one of the data. 
    Bold line shows the outer edge. 
          }
    \label{op}
\end{figure}

We write $T_c(R_{out}) = T_c (1 - R_c/R_{out})$ 
and estimate $R_c$ from the numerical data. 
The results are shown in the table \ref{rc2table}. 
$R_c$ is obtained from the fitting in the figures. 
If the fitting line is $y = \alpha x + \beta$ where $\alpha \simeq -1$, 
we obtain $R_c = e^\beta$. 
It can be seen that $R_c$ is not close to $R_{c2}$ which is 
estimated from Eq. (\ref{rc2}). 
We found that $R_c$ is rather close to $R_{c3}$ which corresponds to 
the third critical field $H_{c3}$ of type-II superconductors. 
According to Ref. \onlinecite{Tinkham}, 
$R_{c2}$ and $R_{c3}$ are related by $R^{\rm th}_{c3} = 0.59 R^{\rm th}_{c2}$. 
$H_{c3}$ of a type -II superconductor 
gives the critical field of surface nucleation 
of superconducting order in a magnetic field. 
Therefore $R_c$ in a ring CDW may also correspond to the 
same phenomenon. 
This is confirmed by examining the order parameter which 
develops at the transition temperature. 
In Fig. \ref{op}, we have shown the order parameter 
at the critical temperature in case of Fig. \ref{tc_thick} (b). 
Even though the radius is changed, 
CDW order starts developing always 
from the outer edge of the sample and 
it tells us that the CDW transition is of surface nucleation type. 

From the above argument we obtain a rough picture of 
CDW transition in the ring crystals: 
At $T=T_{c3} = T_c (1 - R_{c3}/R)$, 
CDW order starts nucleation at the outer surface of the sample and, 
at $T=T_{c2} = T_c (1 - R_{c2}/R)$, CDW order prevails in 
the entire sample. 

\begin{table}
\begin{center}
\begin{tabularx}{80mm}{|C|C|C|C|C|C|C|C|}
\hline
Fig. &
W & $\xi_0/d$ & $k_F/d$ & $\gamma$ & $R_{c2}^{\rm th}/d$ & $R_{c3}^{\rm th}/d$ & $R_{c}/d$
\\ \hline
\ref{tc_thick} (a) & 
$2 \times R$ & $5$ &  $2$ & 0.1 & 10.0 & 5.90 & 5.46 
\\ \hline
\ref{tc_thick} (b) & 
$2 \times R$ & $20$ &  $2$ & 0.1 & 160 & 94.4 & 87.3
\\ \hline
\ref{tc_thin} (a) & 
$R/5$ & $5$ &  $2$ & 0.1 & 10.0 & 5.90 & 5.96
\\ \hline
\ref{tc_thin} (a) & 
$R/5$ & $20$ &  $2$ & 0.1 &  160 & 94.4 & 95.7
\\ \hline
\end{tabularx}
\end{center}
\caption{$R^{\rm th}_{c2}$ is obtained from Eq. (\ref{rc2}). 
$R^{\rm th}_{c3} = 0.59 R^{\rm th}_{c2}$ (see text). 
$R_c$ is obtained from the fitting in Figs. \ref{tc_thick} and \ref{tc_thin}. }
\label{rc2table}
\end{table}

\section{Discussion}

Let us compare our theory with the experimental results 
obtained by Tsuneta et al.\cite{Tsuneta2}

Tsuneta et al. found that the transition temperature 
estimated from the anomalies in resistivity 
does not change significantly between whisker and ring crystals, 
although significant decrease is observed in 
\lq\lq figure-eight\rq\rq\, (twisted ring) cryslats. 
As for the ring crystals, their data are in good agreement 
with our result, since the radiuses of the ring crystals
used in the experiment may be mostly above 
$10\mu$m and then, 
according to our theory, the suppression of 
transition temperature is about  
$T_c R_{c2}/R \sim 0.001\times T_c \sim 0.1$K.
This is smaller than the sample-to-sample fluctuations. 
In contrast to this, in figure-eight crystals, 
the transition temperature is suppressed 
about 1K. 
This significant suppression compared to rings is 
worth noting. 
Although detailed effects of the twisting is 
not clear at this stage, 
we consider that such a large suppression of $T_c$ due 
to geometrical frustration 
is unexpected from the naive intuition based on the 
present analysis. 
The explanation of this phenomenon needs further 
studies, both from theoretical and experimental sides. 

Here we comment on the lower temperature side. 
The free energy, Eqs. (\ref{disc_f1}) and (\ref{disc_f2}), 
is valid only in very vicinity of $T_c$, 
since the effect of electrostatic screening 
(Debye screening) is neglected. 
The dislocations in CDW give rise to the polarization of charge 
arising from $\partial_x \phi_j(x)$, 
and thus undergoes the electrostatic screening due to the quasiparticles 
and the condensate (phasons). 
Since the electrostatic screening is strong in most CDW materials, 
this neglect is serious. 
A free energy which includes the screening effect is 
given, for example, in Ref. \onlinecite{Hayashi-Yoshioka}, 
according to which, the interaction between dislocations 
is not the type of anisotropic GL theory but shows complicated 
dependence on direction and scale. 
In order to clarify the CDW order in ring crystals at much lower 
temperature than $T_c$, a study based on such 
accurate free energy is 
indispensable, which may need 
further numerical studies. 
We leave this for future study. 

It should also be noted that in realistic CDW materials 
the effects of thermal fluctuation are essential. 
As is well known, the real transition temperature of 
CDW is much lower than the mean-field value 
which is estimated from the CDW energy gap. 
This is considered to be due to the renormalization of 
$T_c$ by the thermal fluctuations. 
In order to obtain more reliable results, 
we need to go beyond the mean field approximation, 
For example, numerical simulations as performed 
by Nogawa and Nemoto, 
given in Ref. \onlinecite{Nogawa}, are hopeful. 
Further studies in such direction are desired. 

\section{Summary}

We have derived the free energy which describes the 
CDW order in ring crystals and, based on it, 
calculated the mean-field transition temperature. 
We have discussed several physical properties of ring CDW 
based on the obtained free energy. 

\begin{acknowledgments}
M.H was  financially supported by 
Grants-in-Aid for Scientific Research of Ministry
of Education, Science, and Culture. 
K.K was financially supported by the Sumitomo Foundation. 
\end{acknowledgments}



\begin{thebibliography}{99}
\bibitem{1d}
In this paper we limit ourselves to CDW's in quasi-one-dimensional systems, 
such as NbSe$_3$ and TaS$_3$ and do not consider two-dimensional materials, 
such as NbSe$_2$.
\end{thebibliography}
\end{document}